# SYSTEMS VARIABILITY MODELING: A TEXTUAL MODEL MIXING CLASS AND FEATURE CONCEPTS

Ola Younis[1], Said Ghoul[1], and Mohammad H. Alomari[2]

[1]Bio-inspired Systems Research Laboratory, Philadelphia University, Amman, Jordan
[2]Electrical & Computer Eng. Dept., Applied Science University, Amman, Jordan

*ABSTRACT*

*System's reusability and cost are very important in software product line design area. Developers' goal is to increase system reusability and decreasing cost and efforts for building components from scratch for each software configuration. This can be reached by developing software product line (SPL). To handle SPL engineering process, several approaches with several techniques were developed. One of these approaches is called separated approach. It requires separating the commonalities and variability for system's components to allow configuration selection based on user defined features. Textual notation-based approaches have been used for their formal syntax and semantics to represent system features and implementations. But these approaches are still weak in mixing features (conceptual level) and classes (physical level) that guarantee smooth and automatic configuration generation for software releases. The absence of methodology supporting the mixing process is a real weakness. In this paper, we enhanced SPL's reusability by introducing some meta-features, classified according to their functionalities. As a first consequence, mixing class and feature concepts is supported in a simple way using class interfaces and inherent features for smooth move from feature model to class model. And as a second consequence, the mixing process is supported by a textual design and implementation methodology, mixing class and feature models by combining their concepts in a single language. The supported configuration generation process is simple, coherent, and complete.*

## 1. INTRODUCTION

Designing product lines process has received potential attention recently. This is due to the need of decreasing software product line steps and increasing system reusability. Software Product Line (SPL) is the process of developing products' components from pre-defined core assets rather than developing each component individually [1]. SPL approaches attempt to increase system's productivity by designing a set of products that have many commonalities and shared characteristics, which leads to increasing system's reusability. On the other hand, SPL aims to identify and manage the variations among the products [2]. Product line commonalities and variabilities are composed together in the Domain Space model as feature models and these models form the basic structure for future releases and system variant products [1]. A linked model named Solution Space is connected to the Domain Space to represent the real assets for variability elements associated with some rules to ensure valid selection and consistent system release generation [3]. Several techniques are used to model domain space and solution space. Feature modeling is the most famous technique for this purpose [1, 2]. For modeling solution space, class models are used with some other options like Domain Specific languages (DSL) compilers, generative programs and configuration files [4].

Over the past few years, several research contributions were reported to handle SPL variability process. They can be classified according to SPL's development methodology (requirements, analysis, design, and implementation) or the techniques they used to represent variability (text, graph, or mixed). Approaches that used object-oriented paradigm [5, 6] to model variability described system architecture by package diagrams that used class diagrams. Several approaches [7-11] mix feature models with class models to present software product line engineering process. These approaches designed the variability and commonalities between variants of a product based on features with feature model, and implement these variations in class model. The mixing was done using several techniques like constraints additions [8, 12], relation definition [7, 10] and references links [11]. These approaches defined the way for instantiating objects (configuration) that provides the final product (release) from selecting objects based on fixed features and resolving constraints and relations among them. Approaches supporting SPL requirement and analysis are good for providing general view of systems' needs and characteristics, but, they do not support system functionalities like approaches covering design and implementation steps.
Graphical object-oriented modeling approaches provide clear representation for system hierarchy and components relations. While textual object-oriented approaches give very strong semantic representation for system components and relations, but they are weak to represent the hierarchy relations and structure. Both textual and graphical object–oriented approaches are limited in modeling variability, because of absence of features. Approaches that mix feature and class models encounter insufficient mixing techniques. This does not provide powerful languages that mix system's feature and variability implementation [1].

From the above research context, the following challenges may be stated: (1) Design and implementation approaches are very challenging phases, because they bridge between conceptual and implementation levels. (2) Variability design and implementation methodology are generally missed; their introduction and specification will lead to a great enhancement of SPL. (3) Mixing class and features models through new languages which are so far to be mature, evaluated, and accepted. Conceptual enhancements and practice evaluation will promote these valuables approaches to industrial level. (4) Configuration generating approaches are complex and aiming to generate coherent and complete objects. Ensuring the simplicity, coherence, and completeness of these kinds of objects remain always as open problems.

This paper, propose a Textual Software Product Lines Design Model, mixing class and feature concepts, and aiming to bring significant solution elements to the previous problems, through its specific SPL Methodology: (1) Provide a formal methodology supporting variability design and implementation. It bridges between product lines design model and object oriented implementation model. (2) Provide a new concise and rich textual notation for feature modeling and class modeling. It allows simple and natural new way of mixing feature models and class models using small number of concepts. And (3) allow simple, coherent, and complete configuration generation as simple class instantiation. In the following, we start by the literature review of approaches mixing class and features models, in order to provide evidences motivating our work. The second section introduces our approach, A Textual Software Product Lines Design Model Mixing Class and Feature Concepts, through a new developed methodology supporting variability design and implementation. This approach will be evaluated and compared with others' works in the third section in addition to a conclusion and expected future works.

## 2. SIMILAR WORKS ON MIXING CLASSES AND FEATURES MODELS

Large systems that are composed by huge number of different components cover multiple ideas and variant areas of interests. Thus, each of its components may have more than one possible value to cover. These values came from domain analysis, stockholders' needs, system evolution and so many other cases. The ability of a system to be generalized specialized or customized [2] to perform special needs is called system variability and specified using feature modeling. This section starts with listing feature modeling fundamentals and then presents a review of previous similar works that mix class models and feature models in system variability modeling.

### 2.1. Features modeling fundamentals

Over years of variability modeling, feature modeling using features diagrams was the most popular technique to represent variability in clear and meaningful way [1]. Researches in feature modeling can be classified in three main groups based on the technique they used. Some approaches used pure graphical representation for their feature model's syntax and semantics like ECORE [11] and OOFM [13], and the work reported by Laguna and Marques [4], Razieh et al [14], and Teixeira et al [15]. Other approaches choose to use textual representation for their feature model's syntax and semantics like TVL [16] and FEATUREIDE [17], and the work reported by Arnaud et al [18]. In order to benefit from graphical and textual techniques, some approaches mixed them for representing their feature model. These approaches like CLAFER [9, 12] and RBFEATURES [8].

Graphical feature modeling consists of tree hierarchy that shows the variable feature as the head node and the variant features as children nodes [10]. Designers do not prefer to use graphical representation for more than one reason [16]: firstly, designing feature models using graphical representation is considered a very boring process and does not reflect the real semantic of system components. Secondly, graphical representation is very weak in representing system reasoning process [2]. Finally, graphical notation is still a "research prototype" [16] and can't reaches text notations for representing feature models. Its main concepts are [14]: (1) Meta-Features Model. Previous researches did not mention the Meta-Models Clearly; they mentioned it as features that may contain more than one sub features. (2) Features Meta-Model; this model is predefined and domain independent. It defines different domain features with their relations. (3) Feature Model; Compact model of features diagram and feature constrains. It is an instance of the Features Meta Model. (4) Feature diagram; Graphical representation showing each feature and its relations with its subs. And (5) Feature's configuration; Set of selected features producing a release in SPL. Configuration is permitted with feature model and preserves features' constrains.

Textual feature modeling got rid of all these notations and modeling languages for representing features and their relations. It used simple texts composed by grammars, and propositional formulas [18] to show model structure and implementation.

### 2.2. Models mixing classes and features

Feature models used to design system's variability and communality over its components [12]. Class models capture the implementation part of the products by showing the real values and relations over components' attributes. Thus, mixing both models (feature model and class model) provides the full picture for SPL's components. This section presents a review of the works mixing feature models with class models in two phases: (1) how they mix feature models and class models? And (2) how they instantiate objects (configuration) to create final products?
CLAFER model [12] presents a good approach for mixing class model with feature model based on constraints and inheritance concepts. The feature model was presented as a collection of type definitions and features. The mixing between feature model and class model via constraints is added to class model as attributes and attributes' values. The final model is restricted to one configuration based on the mixed feature. Object instantiation in CLAFER is done by adding constraints to the feature model resulting as constrained feature model. These constraints restrict the feature model to single or dual configuration presenting one or more final product.

Gunther and Sunkle [8] reported feature oriented programming language called RBFEATURES on top of dynamic programming language (ruby). The class model was reported as a first-class entity and named ProductLine. Mixing feature model with class model was done via add-feature method. After creating feature model in RBFEATURES, the ProductLine that is created via configure method and collects number of conceptual features. It is allowed to set specific feature configuration with activate_feature and deactivate_feature operations.

Sarinho and Apolinario [10] presented object-oriented feature model that combined feature models' concepts with object-oriented concepts. They proposed object-oriented feature model (OOFM) profile that is composed by feature model and feature modeling package. Feature classes were reported with object-oriented relationships and resources to provide new level of variability documentation. Feature classes can be declared using feature-class stereotype that creates classes according to designer's intentions. This process composed by several steps starting by feature package creation, followed by OOFM profile mapping and ended by class feature declaration.
Bio-inspired aspect-oriented paradigm was presented by Ghoul [7] to reflect biological principles on the artificial systems. The author presented aspect models as Genomes components and class models that implement them. The mixing was done using relation between feature models and class models. Object instantiation is done by a Weaver that guarantees the consistency over all selected components.

Stephan and Antkiewicz [11] reported ECORE, a class model notation that are presented as feature models. Class model consists of a class compose-by hierarchy. Mapping between feature models and class models was done in both ways: feature to class and class to feature. Object model provides a conceptual view of the final product to give designer basic structure of configuration model.

## 2.3. The presented work

The review of the above and others similar works have mainly revealed the absence of: (1) a specified and formalized methodology supporting systems variability engineering, (2) any modularity of domain features with modeling concepts supporting it, (3) a satisfactory approach mixing class and feature concepts, and (4) a suitable configuration technique. Thus, this paper aims to provide some solutions these stated weaknesses. It proposes a software engineering methodology bridging product lines design models and implementation models for creating object oriented SPL. It introduces and specify some modularity concepts; Meta-Feature Model (as a design pattern that specifies feature's structure), Feature Meta-Model, Feature Model, Product Meta-Model, Product Model. It provides a concise and rich textual notation for feature modeling and class modeling. This feature model may be linked with class model in a way that reflects both models concepts. Finally, a simple, coherent, and complete configuration generation method, as simple class instantiation, is proposed.

## 3. A TEXTUAL MODEL MIXING CLASSES AND FEATURES

This section presents our approach for modeling SPL systems. A textual design methodology is introduced with its supported feature modularity concepts and mixing technique. Graph notations are used only for clarity purposes and not as basic syntax notation which is textual.

International Journal of Computer Science & Information Technology (IJCSIT) Vol 5, No 5, October 2013

### 3.1. A Textual Design Methodology (TDM)

In the following, we introduce the textual SPL design methodology (TDM), its features concepts, its object-oriented concepts, its mixed class and features concepts, its illustration by an example, and finally a conclusion on its specification. The TDM, with graph notations showing its ordered steps for designing variable software, is shown in Figure 1.

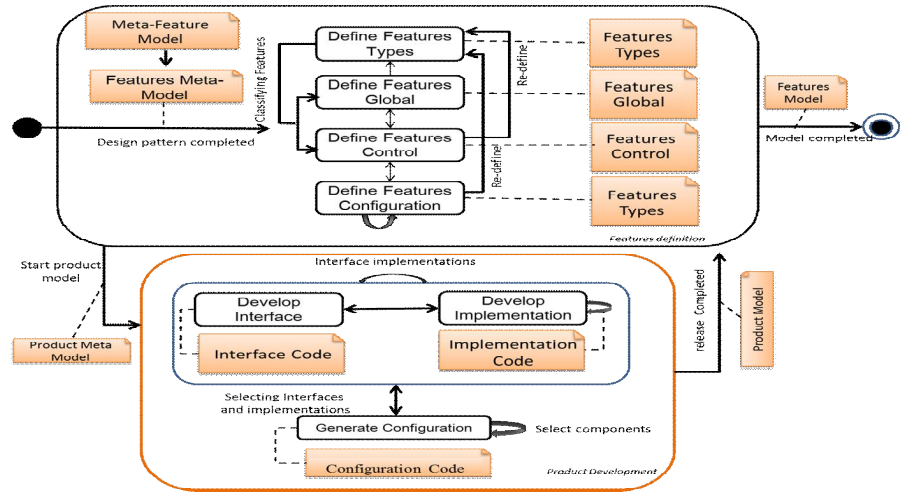

Figure 1 Textual Design Methodology (TDM) mixing class and feature concepts

### 3.2. TDM Features Concepts

Designing steps are based on pre-defined features. A new development will be started by instantiating the Features Meta-Model; which is composed by four feature modules: Features types, Features Global, Features Control, and Features Configuration.

**Meta-Features Models**

It is a predefined design pattern that models all features in TDM. It is the template for features in Features Meta-Model. Its structure is shown in Figure 2. Each feature is composed by a name; an association component to determine its associations with other features, a constraint component that specifies constraints which may affect its relations with others, and finally, a Product features that form its possible values.

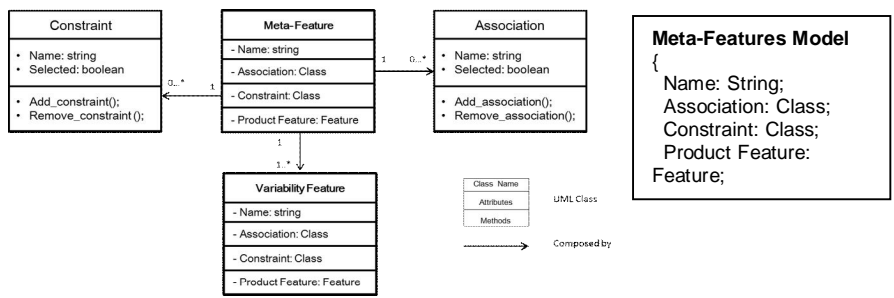

Figure 2 (a) Graphical and (b) textual representations of Meta-Feature Model.

International Journal of Computer Science & Information Technology (IJCSIT) Vol 5, No 5, October 2013

**Features Meta-Model**

It is the input features design pattern to the methodology. It is predefined and based on Meta-Features Model design pattern. It is domain independent, and any feature model (which is domain dependent) is instantiated from it. Figure 3 shows its graphical and textual representation.

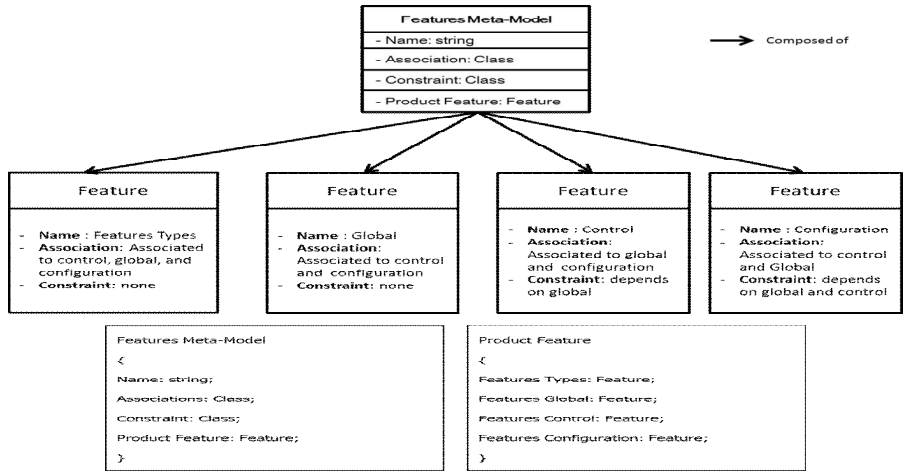

Figure 3 (a) Graphical and (b) textual representations for Features Meta-Model

Below, each feature module is presented separately using graphical and textual notations.
*Features Types.* This feature module captures all features in the system with their possible values. It is composed by Features_Types and Relation_Types. The former represents all systems' features (characteristics); and the later represents all systems' features possible relations. These features and relations will specify the Global, Control, and Configuration features modules. Figure 4 shows graphical and textual representation of Features Types.

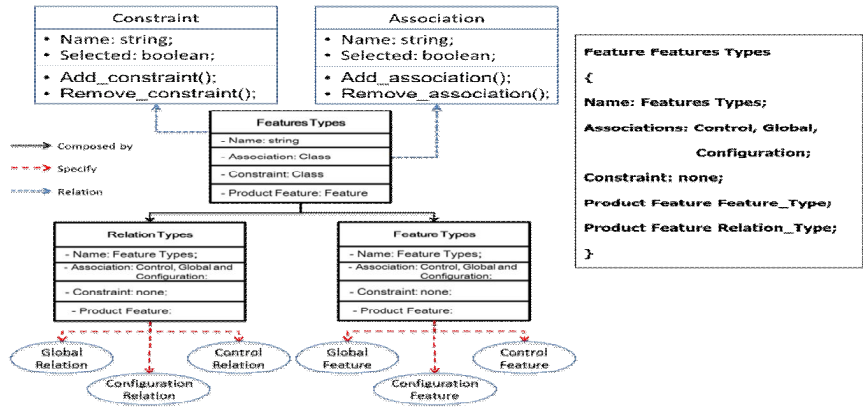

Figure 4 Features Types

*Features Global.* This feature module specifies the *Global* features that will be shared between all system components. A *Global* feature may be relation over components or just feature

International Journal of Computer Science & Information Technology (IJCSIT) Vol 5, No 5, October 2013

(characteristic) that must be applicable everywhere. Figure 5 shows the graphical and textual representations for feature Global.

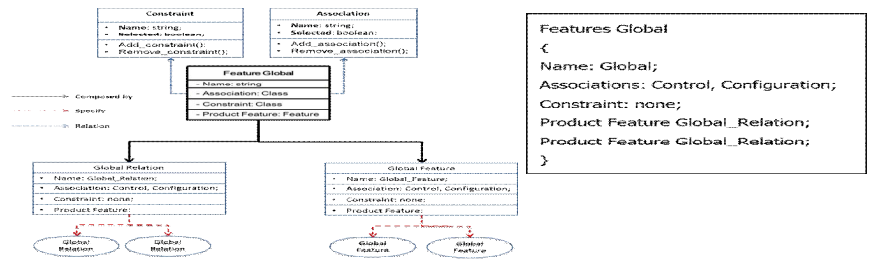

Figure 5 Features Global graphical and textual representations

*Features Control.* This feature module specifies the controls over all systems' components and relations. Any configuration should reserve control's relations to ensure system consistency. This feature module is composed by relations only, and its main goal is to keep systems' components stable and avoid any conflicts. Figure 6 shows graphical and textual representation for feature Control.

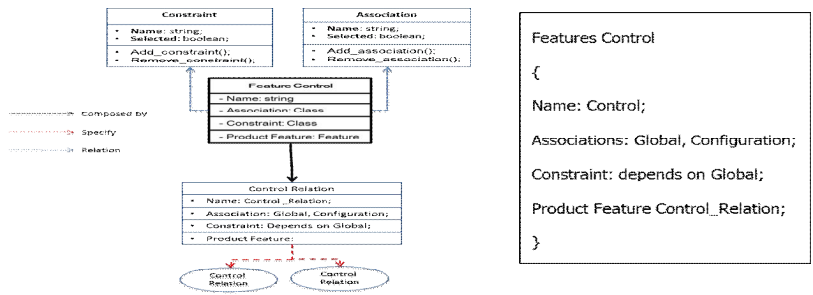

Figure 6 Features Control graphical and textual representations

*Features Configuration*. This feature module specifies required and discarded features for a product configuration (release). Figure 7 shows the graphical and textual representation for Feature Configuration.

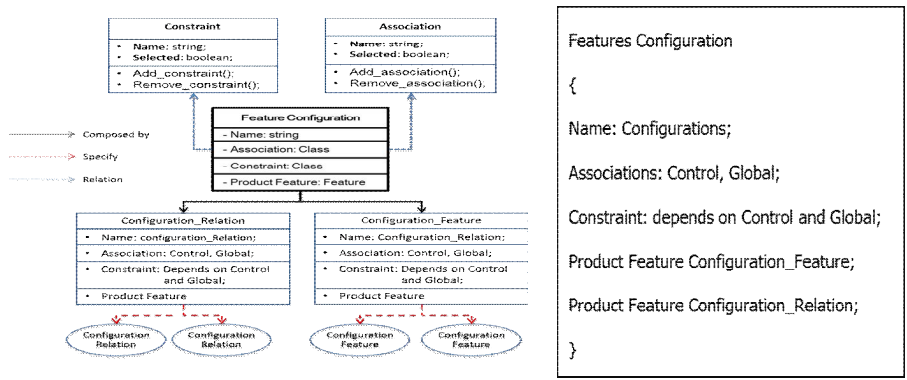

Figure 7. Feature Configuration

Features Types, Global, Control and Configuration together compose the Features Meta-Model in TDM. The second step is creating Feature Model from these Features Meta-Model.

**Features Model**

This is an intermediate model between the conceptual part (Feature Meta-Model) and the physical part (Product Model). In this model, all features and relations in the Features Meta-Model are instantiated for a specific domain. Figure 8 shows instantiation of Features Model from Features Meta-Model. Meaning that each Feature Meta-Model may have one or more instances in its Features Model. Thus, the cardinality relation between them is one to many.

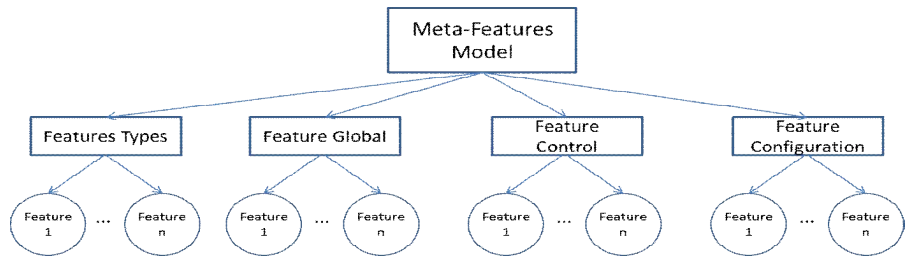

Figure 8 Instantiation of Feature Model from Feature Meta-Model

### 3.3. TDM Object-oriented Concepts

In this section, we report the object-oriented concepts that TDM covers through its Product Meta-Model (Figure 9, Figure 10).

Class Interface specifies services provided by a product component. It includes its provided methods, its attributes (data) and its different implementations' list. Figure 9 shows the graphical and textual representations of Class interface.

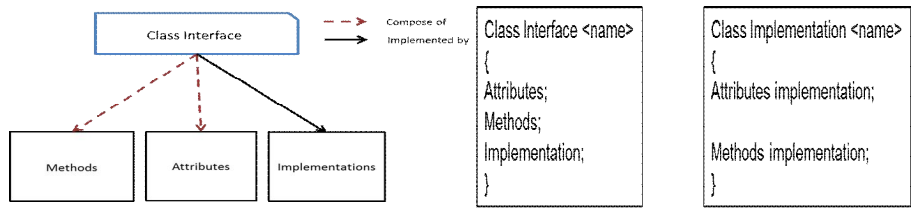

Figure 9 Graphical and textual models for class interface

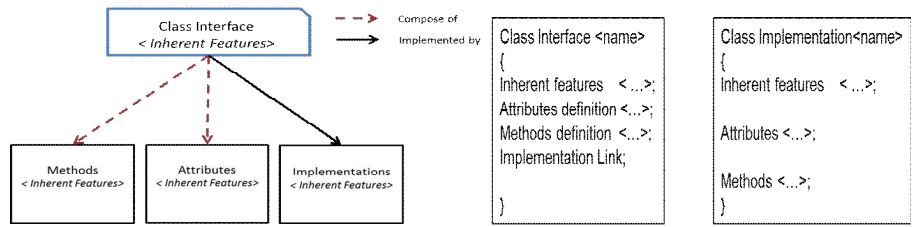

Figure 10 Product Meta-Model (graphical and textual models)

International Journal of Computer Science & Information Technology (IJCSIT) Vol 5, No 5, October 2013

### 3.4. TDM Mixing Class and Features Concepts

This section exposes the mixed class and features concepts that TDM covers through its Product Meta-Model and Product Model.

Product Meta-Model: It is the TDM object-oriented meta-model mixed with features (defined from domain), and inherent features (that are defined for each component based on its properties). It is composed by Interface Meta-Model and Implementation Meta-Model as shown in Figure 10. Each attribute or method can be defined in several ways depending on the features it composes. Each time a new feature is added to an interface component, a new definition should be held.
Product Model: This is the final model. It is composed by class interfaces and their specified attributes, methods, and implementations. Figure 11 shows the graphical and textual representations for this model.

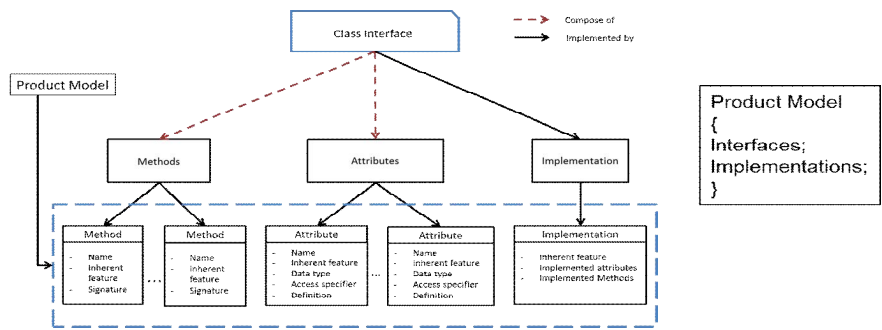

Figure 11 Product Model (graphical and textual models)

### 3.5. A case study

A *Set* is a variable class, having several model versions such as: Static stack, static queue, dynamic stack and dynamic queue. In the following, we present some significant parts of this case study. The complete example is presented in [19].

*Feature Model*: The Feature Model of the set is composed by its *Features Types, Feature Global, Feature control, and Feature Configuration.* Figure 12 shows the features types model of Set. The Figure 13 shows Set *Features Control model*. It is responsible of controlling the relations over model components.

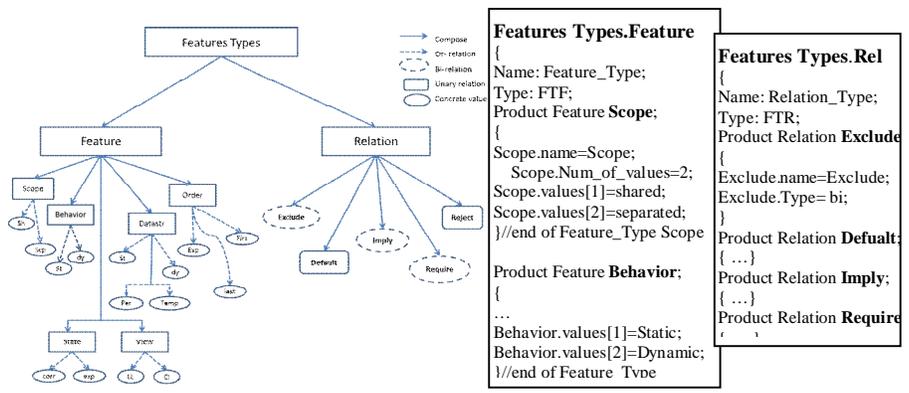

Figure 12 Set Feature Types (graphical and textual models)

International Journal of Computer Science & Information Technology (IJCSIT) Vol 5, No 5, October 2013

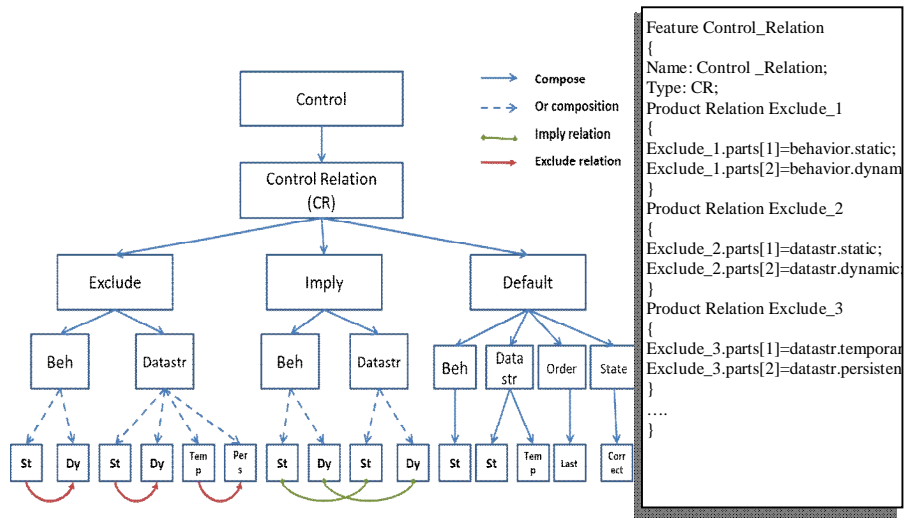

Figure 13 Set's Control Features (graphical and textual models)

*Product Model*: Figure 14 shows the final product model for Set example. The figure specifies the Set interface, Stack sub-interface, Stack implementation, and a Stack configuration.

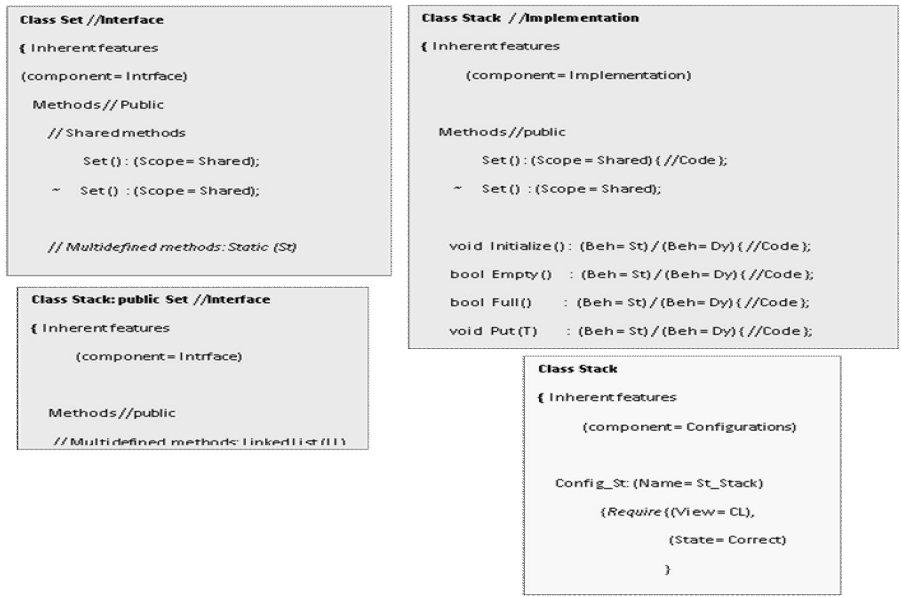

Figure 14 Set Product Model

## 4. IMPLEMENTATION ISSUES, EVALUATION, AND APPLICATION AREAS

### Implementation issues

The implementation environment of this methodology requires a strongly typed Object-Oriented Programming Language (OOPL). The checking process should guarantee the correct association between the Meta-Features model, Features Meta Model, Features Types, Features Global, Features Control, Features Configurations, Product Meta Model, and Product Model. We needed to extend C++ for supporting the concepts of TDM. We are building on extensions that were presented earlier in [7]. These extensions might be processed by any OOPL pre-processor. Configurations can be created, as object instances, according to their Feature models.

### Application areas

Software engineering processes will be strengthened by adding TDM to their feature modeling techniques, since it TDM is more powerful than current conventional approaches in presenting features and classifying them. TDM is highly recommended to be used in any feature modeling area like configuration management, feature-oriented programming, product family engineering and software product lines. A real example of systems that may use TDM in their programming is operating system implementation, multi-agent systems and any system that needs variability.

### Evaluation - TDM Concepts vs. conventional related concepts

In the following, we compare the power of TDM concepts (new or enhancement of old ones) with similar conventional related ones in similar works. . The following table (Table 1) summarizes this comparison.

Table 1 TDM Concepts vs. conventional related ones.

| Concept | Current approaches | TDM approach |
|---|---|---|
| Features Meta-Mode | Conventional approaches like [4, 8, 9, 12] and other research works have described meta-model in term of features that have more than one sub-features as children. | Features Meta-Model is enriched by the features **modularity**: *Features Types, Features Global, Features Control* and *Features Configuration* modules were introduced. The relations between these modules are specified. |
| Features Types | Each feature is defined individually. No support for full declaration for systems' features. | TDM provides **strong typing** of all features (their possible values and relation declaration in the Features Types module). |
| Features Global | Shared features are not separated as a unit, but defined in the feature diagram hierarchy. | Global features are **separately** defined in the Feature Global module |
| Features Control | Relations between features are not separated as a unit, but defined along with features. | Control features are relations that specify coherence of configuration. They are **separated** in the Feature Control module |
| Features Configuration | They are presented along with the feature model. | They are **separated** in Feature Configuration module. |
| Methodology | Conventional approaches are not supported by a specified and formalized methodology | TDM is fully **supported by a methodology** integrating the variability design and implementation |
| Product Interface | Conventional approaches are weak in support component's interfaces. Each class is created based on its configuration characteristics. | TDM supports component interface and implementation increasing the modularity and mixing the features concepts with object-oriented ones |

## 5. CONCLUSION AND PERSPECTIVES

Through our study about feature modeling and SPL engineering, we found that current feature models did not support feature modularization. Linking feature models with class models is still

International Journal of Computer Science & Information Technology (IJCSIT) Vol 5, No 5, October 2013

weak and does not reflects feature model's concepts, and there is a lack in variability design and implementation methodology. We proposed four enhancements. The first was the textual feature design methodology that supports software product line engineering. The second was the modularization of features through four meta-feature models that classify features according to their functionalities. The third was the link between feature model and class model to allow mixing features' concepts with real implementation of classes. And finally, the last was an approach to configuration generation based on pre-selected features. TDM might be extended and developed in future to: (1) define other meta-feature models to capture all software's variability features. (2) Enhance current class model to be more realistic and reflects feature model in uniform and formal way. (3) Enhance the configuration generation to be a smart automated generation. And (4) Design a uniform language mixing features and classes.

## REFERENCES

[1] J.-M. Jézéquel (2012) Model-Driven Engineering for Software Product Lines, Report 24.
[2] S. Marco and D. Sybren (2007) "Classifying variability modeling techniques", Information and Software Technology, Vol. 49, pp. 717-739.
[3] M. Marcilio, B. Moises, and C. Donald, S.P.L.O.T. (2009) "Software product lines online tools", Proceedings of the 24th ACM SIGPLAN conference companion on Object oriented programming systems languages and applications, Orlando, Florida, USA, pp. 761-762.
[4] M. A. Laguna and J. M. Marques (2009), "Feature Diagrams and their Transformations: An Extensible Meta-model", 35th Euromicro Conference on Software Engineering and Advanced Applications, pp. 97-104.
[5] A. Savinov (2012), "Concept-oriented programming: classes and inheritance revisited", 7th International Conference on Software Paradigm Trends, Rome, Italy, pp. 381-387.
[6] T. Sim-Hui (2013), Problems of Inheritance at Java Inner Class, ArXiv e-prints.
[7] S. Ghoul (2011), "Supporting Aspect-Oriented Paradigm by bio-inspired concepts", 4th International Symposium on Innovation in Information & Communication Technology (ISIICT), pp. 63-73.
[8] S. Gunther and S. Sunkle (2012), "rbFeatures: Feature-oriented programming with Ruby", Science of Computer Programming, Vol. 77, No. 3, pp. 152-173.
[9] B. Kacper, C. Krzysztof, and W. Andrzej (2011), "Feature and meta-models in Clafer: mixed, specialized, and coupled", Proceedings of the Third international conference on Software language engineering, Eindhoven, The Netherlands, pp. 102-122.
[10] V. T. Sarinho and A. L. Apolinario (2010), "Combining feature modeling and Object Oriented concepts to manage the software variability", IEEE International Conference on Information Reuse and Integration (IRI), pp. 344-349.
[11] M. Stephan and M. Antkiewicz (2008), "Ecore.fmp: A tool for editing and instantiating class models as feature models", University of Waterloo, Waterloo, Technical Report.
[12] B. Kacper (2010), "Clafer: a unifed language for class and feature modelling", Generative Software Development Lab, Technical report.
[13] V. T. Sarinho, A. L. Apolinario, and E. S. de Almeida (2012), "OOFM - A feature modeling approach to implement MPLs and DSPLs", IEEE 13th International Conference on Information Reuse and Integration (IRI), pp. 740-742.
[14] B. Razieh, N. Shiva, Y. Tao, G. Arnaud, and B. Lionel (2012), "Model-based automated and guided configuration of embedded software systems", Proceedings of the 8th European conference on Modeling Foundations and Applications, Kgs. Lyngby, Denmark, pp. 226-243.
[15] L. Teixeira, P. Borba, and R. Gheyi (2011), "Safe Composition of Configuration Knowledge-Based Software Product Lines", 25th Brazilian Symposium on Software Engineering (SBES), pp. 263-272.
[16] A. Classen, Q. Boucher, and P. Heymans (2011), "A text-based approach to feature modelling: Syntax and semantics of TVL", Science of Computer Programming, Vol. 76, No. 12, pp. 1130-1143.
[17] T. Thaum, C. Kustner, F. Benduhn, J. Meinicke, G. Saake, and T. Leich (2012), "FeatureIDE: An extensible framework for feature-oriented software development", Science of Computer Programming, In Press, Corrected Proof, doi: 10.1016/j.scico.2012.06.002.

[18] H. Arnaud, B. Quentin, H. Herman, Rapha, M. l, and H. Patrick (2011), "Evaluating a textual feature modeling language: four industrial case studies", Proceedings of the third international conference on Software language engineering, Eindhoven, The Netherlands, pp. 337-356.
[19] O. Younes (2013), A Textual Software Product Lines Design Model By Mixing Class and Feature Concepts, MSc Thesis, Philadelphia University.

**Authors**

**Ola A. Younis** received her B.S. degree in Computer Science from Jordan Uneversity For Science and Technology (JUST) in 2010 and the M.S. degree in Computer Science from Philadelphia University, Jordan 2013. She has been working as an Assistant lecturer from 2011 to 2012.

**Said Ghoul** has obtained his Master and Ph.D. Degrees, in Software Engineering, from University of Grenoble, France. His research interest includes Bio – inspired systems modeling and supporting concepts and methodologies. He is actually Full Professor at Philadelphia University - Faculty of Information Technology, Jordan.

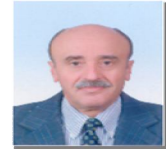

**Mohammad Alomari** is an Assistant Professor at Applied Science University, Jordan. He received his BSc and MSc degrees in Electrical Engineering (Communications and Electronics) from Jordan University of Science and Technology, Jordan, in 2005 and 2006, respectively and PhD in Computer Science and Engineering from the University of Bradford, UK in 2010. His research interests include Computer Vision, Space Weather, Brain Computer Interfaces, and Digital Signal Processing.

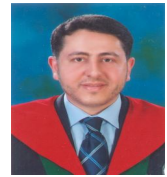